# Magnetically tunable telecom emission from $Er^{3+}$ ions in layered $WS_2$

**Guadalupe García-Arellano[1], Gabriel I. López-Morales[4], Johannes Flick[1,2,3], Cyrus E. Dreyer[3,4], and Carlos A. Meriles[1,2,†]**

[1]*Department of Physics, CUNY-City College of New York, New York, NY 10031, USA.*
[2]*CUNY-Graduate Center, New York, NY 10016, USA.*
[3]*Center for Computational Quantum Physics, Flatiron Institute, 162 5th Avenue, New York, New York 10010, USA.*
[4]*Department of Physics and Astronomy, Stony Brook University, Stony Brook, New York, 11794-3800, USA.*

†E-mail: cmeriles@ccny.cuny.edu

Erbium ions ($Er^{3+}$) provide a telecom-band optical transition with strong magnetic-dipole character, making them attractive for quantum communication and spin–photon interfaces. Identifying host environments that combine low decoherence with photonic compatibility, however, remains a central challenge. Here we investigate $Er^{3+}$ emission in tungsten disulfide ($WS_2$) flakes, a layered host offering low nuclear-spin density and narrow telecom emission. Using time- and polarization-resolved photoluminescence under modest magnetic fields (< 0.2 T), we observe pronounced dimming, lifetime extension, and rotation of the emission dipole when the field has an out-of-plane component, whereas in-plane fields produce little change. Effective model calculations of $Er^{3+}$ in monolayer $WS_2$ parametrized from density functional theory indicate that these effects arise primarily from Zeeman-induced mixing of near-degenerate crystal-field sublevels, which modifies the magnitude and orientation of the optical transition dipole moments. Comparative measurements in flakes of different thickness and numerical estimates of the local density of optical states further suggest a secondary contribution from dipole coupling to the anisotropic photonic environment of thin $WS_2$ layers. These findings identify layered $WS_2$ as a platform where magnetic fields can tune telecom emission through an interplay of crystal-field physics and anisotropic photonic coupling.

## I. INTRODUCTION

Erbium ions ($Er^{3+}$) have long played a central role in optical technologies because their intra-4*f* transition near 1.54 μm lies squarely within the telecommunications C-band, the spectral window of lowest loss in silica optical fibers[1]. In most conventional hosts, this transition is dominated by magnetic dipole (MD) character with additional electric dipole (ED) components[2,3]. The MD contribution is of particular interest: (i) it renders the emission relatively insensitive to electric-field noise, (ii) enables coupling to the magnetic component of light, and (iii) provides a natural interface between long-lived spin states and photons at telecom wavelengths[4]. These properties place $Er^{3+}$ among the most promising solid-state platforms for quantum communication, quantum memories, and spin–photon interfacing[5].

Considerable effort is currently devoted to identifying host materials that can optimize the performance of $Er^{3+}$ for quantum information applications[6]. A central requirement is a crystalline environment with minimal nuclear spin noise, which otherwise limits coherence. Cerium oxide, for example, has been investigated as an attractive host owing to its low concentration of spin-active nuclei[7]. More recently, emission from $Er^{3+}$ incorporated into tungsten disulfide ($WS_2$) flakes has been demonstrated[8]. $WS_2$ provides several advantages: it offers a low nuclear spin environment, it hosts centrosymmetric sites where $Er^{3+}$ likely substitutes tungsten atoms, and the comparable ionic radii of erbium and tungsten minimize strain. As a result, narrow telecom-band emission lines have been observed even in relatively dense ensembles, indicative of very low inhomogeneous broadening. Beyond the spectral quality, 2D materials such as $WS_2$ are compelling because they enable intimate near-field coupling between the emitter and engineered photonic structures, thereby decoupling the challenges of emitter placement, host optimization, and photonic integration.

In crystalline garnets and oxides, magnetic fields primarily induce Zeeman splittings and modest spectral shifts, but the radiative lifetime remains largely unaffected by orientation[9]. In contrast, layered transition-metal dichalcogenides (TMDs) such as $WS_2$ present a highly anisotropic optical environment. They are uniaxial dielectrics with a strong in-plane/out-of-plane index contrast, and when prepared as finite-thickness flakes they act as subwavelength dielectric slabs that support guided and leaky modes[10]. As a result, the local density of optical states (LDOS) sampled by an emitter becomes strongly orientation-dependent, creating opportunities to control emission with external fields in ways that bulk hosts cannot provide.

Here we investigate the emission of $Er^{3+}$ ions coupled to $WS_2$ flakes under applied magnetic fields. We find that modest fields (< 0.2 T) applied normal to the flake, or at 45°, markedly dim the $Er^{3+}$ photoluminescence and lengthen the excited-state lifetime, whereas parallel fields produce little change. Polarization-resolved measurements further reveal a rotation of the emission dipole with out-of-plane fields. Effective model calculations of $Er^{3+}$ in monolayer $WS_2$ parametrized using density functional theory (DFT) point to an "atomic" effect arising from magnetic-field-induced changes in the transition dipole moments (TDM) associated with near-degenerate crystal field (CF) sublevels in the $Er^{3+}$ ground and first excited manifolds. The underlying mechanisms involve an interplay between electric- and magnetic-dipole components of the optical TDMs of the CF Kramers doublets in the presence of Zeeman couplings leading to longer lifetimes and rotated emission dipoles. We also show that the angular dependence of the field-induced changes will depend on the magneto-crystalline anisotropy of the CF doublets involved in the optical transition. Lastly, numerical estimates of the LDOS and comparative measurements in flakes of different thickness suggest that the reoriented dipole can couple differently to the anisotropic photonic environment of thin $WS_2$ layers, providing a secondary contribution to the observed lifetime and polarization changes.

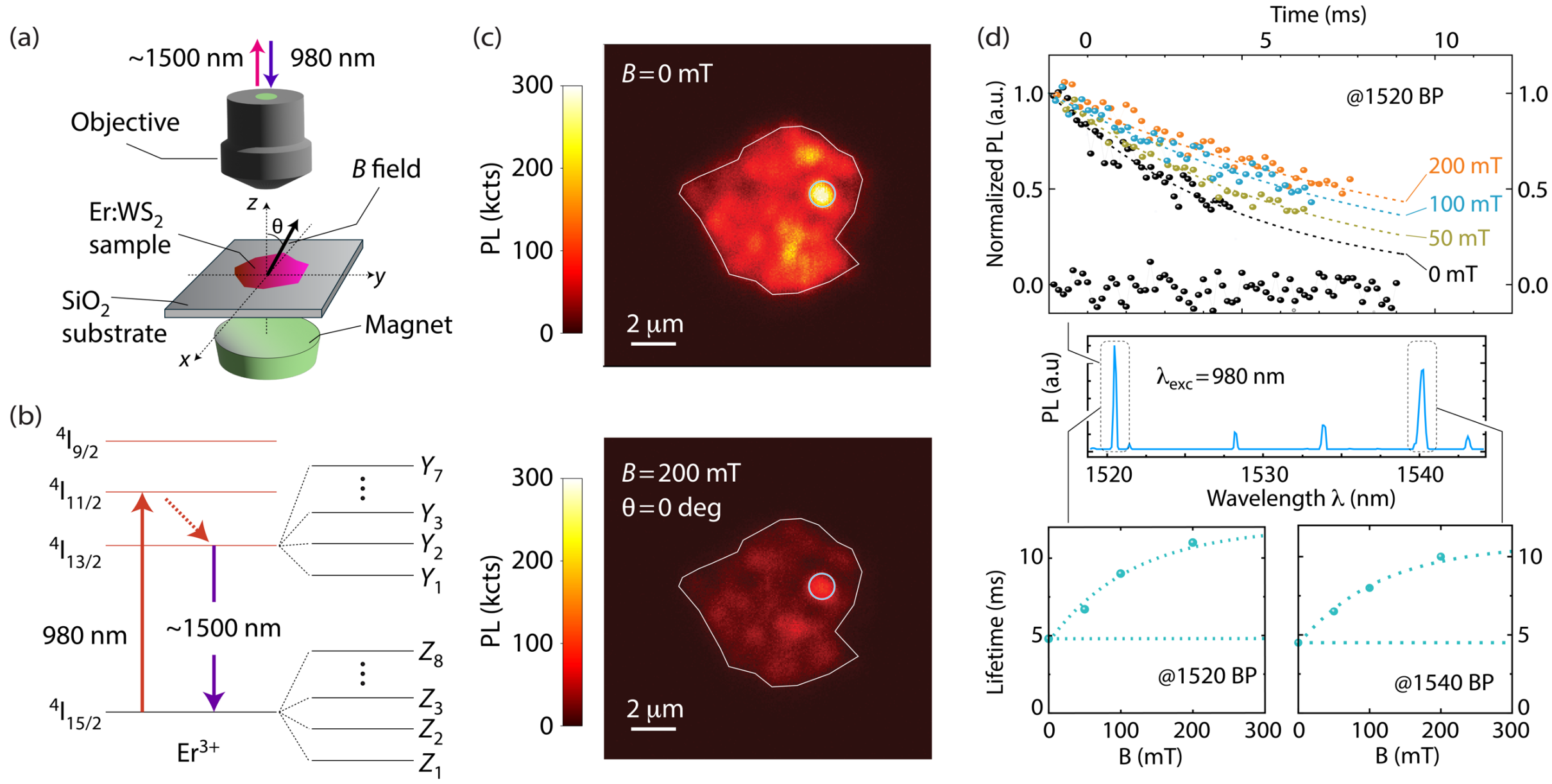

**Figure 1. Magnetic response of $Er^{3+}$ emission in $WS_2$.** (a) Schematic of the experimental setup: an $Er:WS_2$ flake on a $SiO_2$ substrate is excited at 980 nm and imaged through a microscope objective while a magnetic field is applied at an angle $\theta$ relative to the sample normal. (b) Simplified $Er^{3+}$ energy-level diagram showing 980 nm excitation into the $^4I_{11/2}$ manifold and radiative decay near 1500 nm from the $^4I_{13/2} \rightarrow {}^4I_{15/2}$ transition, which we collect selectively with the help of passband filters (not shown). (c) PL maps recorded under zero field (top) and 200 mT normal to the sample plane (bottom), illustrating pronounced emission dimming under out-of-plane field. Here, photon collection spans a 10 nm window centered at 1540 nm. (d) (Middle) Emission spectrum under 980 nm excitation as collected from the circled site in (c) at zero magnetic field. (Top) Time-resolved PL amplitude at 1520 nm (left-hand rounded rectangle in the optical spectrum) at different magnetic fields following pulsed 980 nm excitation; dashed lines indicate single-exponential fits and the bottom trace is a reference obtained in the absence of illumination. (Bottom) Extracted lifetimes from spectrally selective lifetime measurement at 1520 and 1540 nm (left and right plots, respectively) as a function of the applied magnetic field. All experiments at room temperature; a.u.: arbitrary units.

## II. RESULTS AND DISCUSSION

The $WS_2$ flakes we studied in our experiments were exfoliated from a high-purity crystal (2D Semiconductors) and transferred onto a $SiO_2$ substrate, followed by erbium ion implantation (75 keV, $10^{14}$ ions/cm$^2$) and thermal annealing (400 °C for 1 hour)[8]. Figure 1a shows the experimental geometry: We employ a customized confocal microscope operating under ambient conditions, with 980 nm excitation and collection of photoluminescence (PL) in the 1500–1600 nm spectral window, which we detect with a superconducting wire photodetector. We make use of precision stages to control the position and orientation of a permanent magnet yielding a field of variable amplitude $B$ and angle $\theta$ relative to the surface normal (see Appendix A for additional details).

One of the advantages of working with rare-earth emitters is that their optically active inner 4*f* electrons are well shielded from the surrounding crystal environment. As a result, the energy level structure remains nearly unchanged across different host materials[6,11], which greatly facilitates both modeling and experimental implementation. The energy diagram in Fig. 1b summarizes the optical transitions involved: The ground-state $^4I_{15/2}$, intermediate $^4I_{11/2}$, and excited $^4I_{13/2}$ manifolds shown on the left arise from spin–orbit splitting of the $4f^{11}$ configuration. The degeneracy in each manifold is partially lifted by the local CF into sets of Kramers doublets. In the presence of a magnetic field, these doublets experience Zeeman splitting and mixing, which, as we show below, are central to our key observations. Throughout our experiments, we excite the $^4I_{11/2}$ manifold at 980 nm and focus only on the $^4I_{13/2} \rightarrow {}^4I_{15/2}$ emission near 1550 nm, spectrally isolated using bandpass filters.

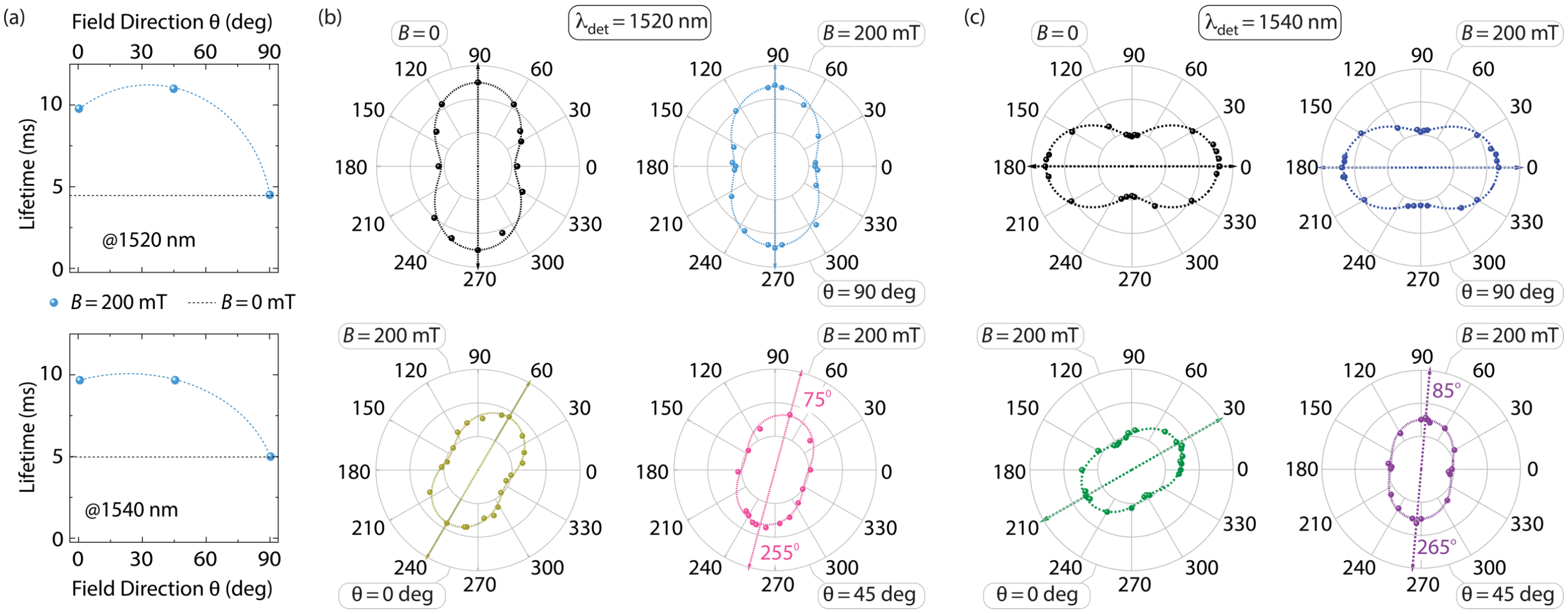

**Figure 2. Angular dependence of lifetime and polarization-resolved emission.** (a) Excited-state lifetime versus magnetic field orientation $\theta$ for detection at 1520 nm (top) and 1540 nm (bottom). Dashed black lines indicate zero-field lifetimes while light blue lines are guides to the eye for 200 mT data. (b) Polar plots of emission linear polarization at 1520 nm under zero field (left) and at 200 mT with the field oriented along $\theta$ = 0°, 45°, and 90°. Field application is accompanied by a rotation of the main polarization axis. (c) Same as in (b) but the emission peak at 1540 nm yielding a similar response.

Figures 1c and 1d illustrate the effect of magnetic fields on the $Er^{3+}$ emission from an implanted flake. At zero field, the PL map (Fig. 1c, top) shows bright emission distributed across the flake. In the presence of a 200 mT field normal to the sample plane, we observe an overall decrease in fluorescence to less than one third of the field-free value (Fig. 1c, bottom). To investigate the origin of this dimming, we recorded spectra and time-resolved emission from selected sites, such as the circled region in Fig. 1c. The spectrum at zero field (Fig. 1d, middle) exhibits narrow emission features centered near 1540 nm, consistent with radiative decay from the $^4I_{13/2}$ manifold. We did not observe any measurable spectral shifts or splitting in the presence of magnetic field (limited by our setup to 200 mT), likely because the ~30 GHz inhomogeneous linewidth masks the expected Zeeman splitting. As shown in the top plot of Fig. 1d, however, time-resolved traces at 1540 nm show that the applied field substantially extends the fluorescence decay time by up to a factor ~2.5. This finding suggests the observed dimming in the PL image of Fig. 1c stems from an extended excited-state lifetime rather than from additional nonradiative channels. In fact, plots of the characteristic decay time reveal similar responses at 1520 and 1540 nm (respectively, left and right graphs at the bottom of Fig. 1d), featuring monotonic growth with the applied magnetic field and saturation at about 12 ms.

The effects highlighted in Fig. 1 for normal-incidence magnetic fields depend strongly on the field orientation. In fact, in-plane magnetic fields ($\theta$ = 90°) produce virtually no changes in emission intensity or lifetime, in contrast to the pronounced dimming observed for out-of-plane fields. Figure 2a quantifies such angular dependence by plotting the excited-state lifetime versus $\theta$. For both the 1520 nm and 1540 nm transitions the response is non-monotonic, with the largest contrast appearing when the field is tilted by about 45° relative to the sample normal. This behavior points to a more complex Zeeman-induced mixing of the Kramers sublevels than what would be expected from a simple projection of the field along one axis.

Beyond its effect on the lifetime, the magnetic field also reshapes the emission polarization, which at zero field is already partly linear. Figure 2b shows the case of the 1520 nm line. At zero field the emission is linearly polarized along a fixed axis, and a 200 mT field oriented in-plane produces virtually no change in orientation, nor in degree of polarization. As the field tilts away from the plane, however, the emission axis rotates, with the largest reorientation observed for out-of-plane magnetic fields. In this case, the total change in polarization direction reaches ~30°. Figure 2c highlights the corresponding behavior at 1540 nm. Here, the emission is also linearly polarized at $B$ = 0, but along an axis nearly orthogonal to that of the 1520 nm line. Again, under $B$ = 200 mT the emission axis rotates, but with two important differences: (i) the sense of rotation reverses relative to Fig. 2b and (ii) the amplitude of such change is much larger, approaching ~90°. In addition, the angular dependence is non-monotonic; the maximum change occurs at a 45° field orientation before decreasing again, mirroring the lifetime dependence

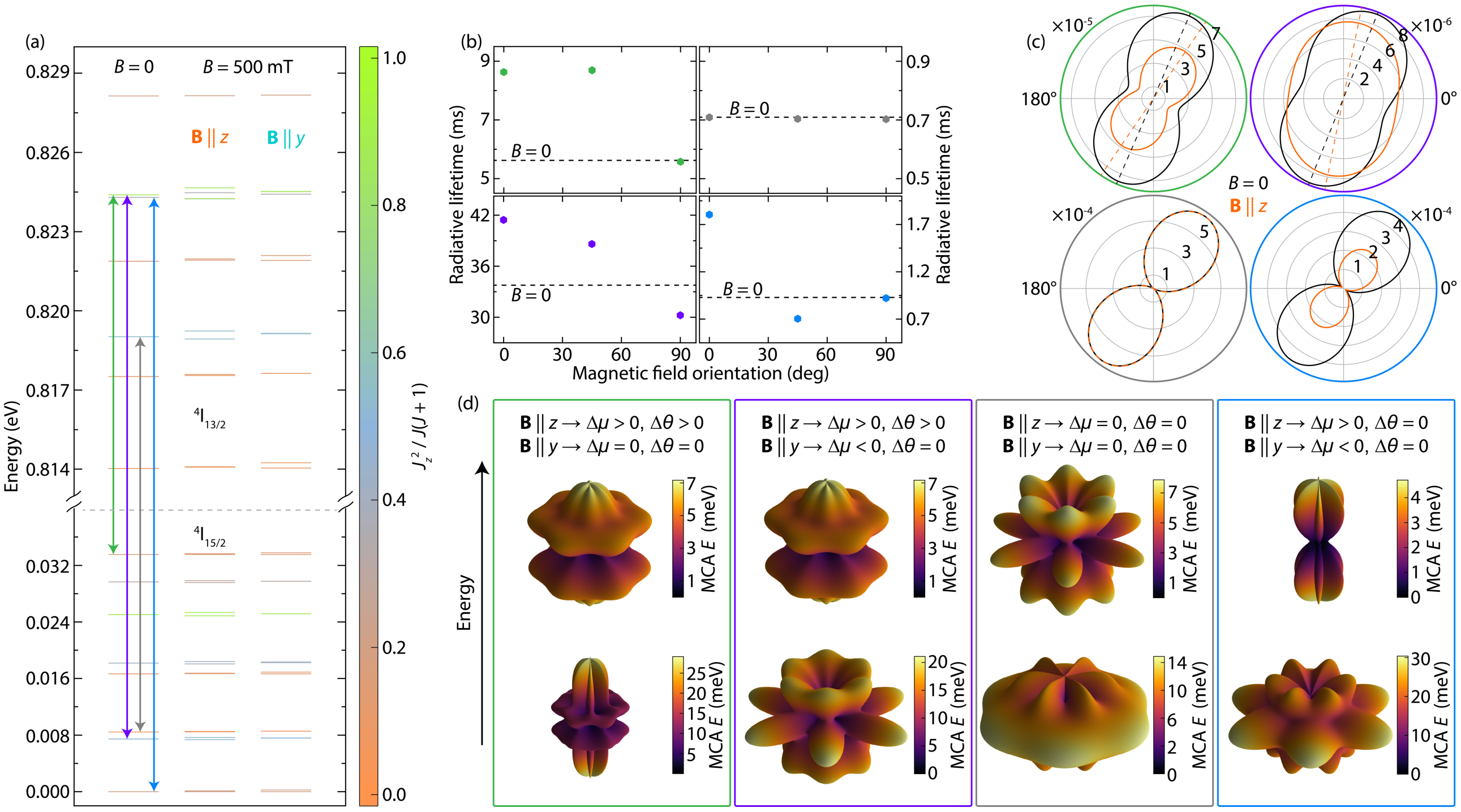

**Figure 3. First-principles modeling of the $Er_W^-$ in monolayer $WS_2$ under external magnetic fields.** (a) Many-body electronic spectra of $Er_W^-$ in the absence of magnetic fields (left), and for $B$ = 500 mT along the $z$ axis (center), and along the $y$ axis (right). A dashed horizontal line is included to separate the ground- and excited-state manifolds. The color scale represents the fraction of the many-body state spread along $z$, while vertical arrows indicate representative transitions. (b) Radiative lifetimes for select transitions, as a function of magnetic field orientation. Each scatter plot in (b) is color-coded with a vertical dashed line in (a) to indicate the corresponding transition; dashed horizontal lines denote lifetimes in the absence of magnetic field. (c) Polarization plots for the same transitions presented in (b) for $B = 0$ and $\boldsymbol{B} \parallel z$ (color-coded for clarity), in units of $e^2 a_0^2$ (where $e$ denotes the elementary charge, and $a_0$ is the Bohr radius). (d) Adiabatic magneto-crystalline anisotropy (MCA) energy surfaces for doublet pairs involved in each of the color-coded transitions. All MCA surfaces share the same (normalized) spatial scale. A summary of the magnetic field effects (i.e., changes in TDM $\Delta\mu$ and orientation $\Delta\theta$) on each transition is also included for clarity.

shown in Fig. 2a.

To elucidate the origin of the field-induced dimming, we compare PL intensity and lifetime measurements. sing the brightest site on the flake as a reference (circle in Fig. 1c), the PL intensity under 10 mW excitation decreases from 250 kcounts/s at zero field to 135 kcounts/s at 200 mT, while the lifetime increases from 4.5 ms to 11 ms. Modeling of the excitation–emission response as a function of the applied laser power (Appendix B), shows that these changes are consistent with a suppression of the radiative rate by a factor of ~2.7. Further, from this analysis and assuming the non-radiative rate remains unchanged in the presence of a magnetic field, we estimate a quantum yield of ~93 % for $B$ = 0 and ~83 % at $B$ = 200 mT. Compared to many oxides or glasses — where the non-radiative decay rate is significantly greater than the radiative rate — this response strongly indicates low crystal inhomogeneity and minimal defect-assisted quenching, in line with the narrow inhomogeneous linewidth we find in the emission spectra[8].

To identify the microscopic mechanisms behind these observations, we employ a semi-empirical correlated model based on density-functional theory (DFT) and quantum embedding of a single $Er^{3+}$ impurity in monolayer $WS_2$[8]. We focus on the negatively charged substitutional defect configuration ($Er_W^-$), where erbium sits at a tungsten site. Note that despite the overall negative charge of the defect complex, the erbium ion itself remains in the trivalent ($Er^{3+}$) oxidation state, with the excess charge localized in nearby dangling-bond and lattice states[12]. In our model, the CF is extracted from DFT calculations via wannierization of the Er 4$f$ states; MD contributions are evaluated assuming ideal 4$f$ states (see Appendix C for a fuller description).

The left column in Fig. 3a shows the calculated CF levels in the absence of an external field. The color scale represents the fractional axial projection of the total angular momentum, $\langle J_z^2\rangle/\langle J^2\rangle$, where $z$ denotes the local CF axis of the Er site, expected to align approximately with the layer normal for substitutional Er. Large values correspond to states with predominantly axial (Ising-like) character and strong $g_z$ factors, whereas smaller values indicate more planar character with stronger transverse response. The broad distribution of $\langle J_z^2\rangle/\langle J^2\rangle$ values therefore implies substantial variation in magnetic anisotropy among the CF doublets. We also find that the CF splitting of the $^4I_{15/2}$ manifold exceeds that of $^4I_{13/2}$, in agreement with previous studies of $Er^{3+}$ ions in low-symmetry CFs[6,11].

A direct one-to-one assignment of the experimental emission lines to specific transitions between individual CF doublets is challenging, in part because the population distribution within the $^4I_{13/2}$ manifold following 980-nm excitation is not known. Instead, we focus on determining the possible mechanisms of the experimentally observed changes with magnetic field. Under the present conditions, we expect the Zeeman splittings to be smaller than the CF, and not observable under the 1–2 nm experimental resolution of our spectrometer. It is thus puzzling how such small changes could have such a significant effect on the optical properties. However, our calculations indicate that several Kramers doublets within both the $^4I_{15/2}$ and $^4I_{13/2}$ manifolds are separated by energies comparable to or smaller than the Zeeman shifts. We find that such near-degeneracies are significant in that they enhance the susceptibility of the optical TDMs to Zeeman mixing, i.e., moderate magnetic fields can produce large dipole modifications even though the Zeeman energy remains much smaller than the overall CF splitting.

To illustrate these dynamics, we focus on a set of representative transitions (vertical lines in Fig. 3a) and determine in each case the radiative lifetimes as derived from incoherent sums of the ED and MD contributions in the presence of a 500 mT external field of variable orientation (Fig. 3b). Interestingly, the response corresponding to the green solid line in Fig. 3a, which involves a near-degenerate set of doublets in the $^4I_{13/2}$ manifold, has a behavior closely resemblant of that seen in experiments, i.e., a significant increase in lifetime selective to an out-of-plane field. The purple and blue lines (which involve the same $^4I_{13/2}$ state) are qualitatively similar, although with slightly different angular dependance. Notably, however, the peak indicated by the grayish line, which does not involve such near-degenerate states, shows very little change in lifetime with magnetic field.

In Fig. 3c, we plot the 2D projection of the radiation patterns for the representative cases identified in Figs. 3a and 3b. Indeed, we find that some of the transitions also feature magnetic-field-induced rotations of the TDM polarization (green/purple cases in the top panel of Fig. 3c), while others remain unchanged by the field (grey/blue cases in the bottom panel of Fig. 3c). Overall, these results reveal that moderate magnetic fields can simultaneously alter the radiative lifetimes and reorient TDMs, qualitatively in line with experiment. Importantly, significant changes are only present if both ED and MD contributions are included for each optical transition, indicating that the underlying mechanism is the interplay between the electric and magnetic components of the optical TDMs.

The angular dependence of the magnetic-field response reflects the magnetocrystalline anisotropy (MCA) of the CF doublets that form the optical transition. In Fig. 3d we show MCA energy surfaces for the Kramers doublets involved in the representative transitions in Figs. 3a-c. The radial and color scales of the surface indicate the energy for the magnetic moment to point in a certain direction. Depending on the doublet, the MCA can be either easy plane (*xy*) or easy axis (*z*). The energy barriers within the $^4I_{13/2}$ manifold are up to six times smaller (2.5× on average) than those associated with doublets across the $^4I_{15/2}$ manifold indicating that magnetic moments in the excited-state are more susceptible to magnetic field-induced reorientation.

By connecting the results across Figs. 3a-d, we can conclude that transitions involving *at least* one easy-plane Kramers doublet are the most susceptible to field-induced spectral changes for out-of-plane fields. Indeed, the grey transition which is basically insensitive to the field involves two easy axis doublets. On the other hand, perturbation theory indicates — to first order — that near-degeneracies can strongly enhance Zeeman-induced state mixing when finite Zeeman coupling matrix elements are present. Provided that the near-degenerate doublets possess non-parallel TDMs, this mixing can result in effective Zeeman-induced "rotations" of the measured TDMs for optical transitions involving the corresponding doublets. In fact, we find that slightly different forms of our CF Hamiltonian yield qualitatively identical results in terms of Zeeman-induced effects and their polar anisotropy, provided a few CF doublets within the excited-state manifold feature degeneracies comparable to the net Zeeman energy (see Appendix C).

While the monolayer DFT model provides a useful framework to capture some of the key ingredients in our observations — namely extended lifetimes, and reorientation of the TDMs — it also has important limitations. First, by treating an isolated $WS_2$ monolayer in vacuum, the model neglects substrate effects and the asymmetric dielectric environment present in experiment. Second, the finite-thickness flakes studied here (200–400 nm) support guided and leaky optical modes that strongly modify the LDOS, whereas the monolayer model only reproduces the intrinsic in-plane/out-of-plane dielectric anisotropy. Third, the calculations do not include radiative coupling pathways or Purcell-type enhancements arising from slab resonances, which can shift both the angular dependence and magnitude of lifetime changes. These considerations suggest that while the monolayer DFT model can reproduce the qualitative features of dipole reorientation under applied fields, it may fall short of fully capturing the underlying dynamics in our experiments.

To assess whether the field-dependent changes are also influenced by photonic effects, we examined Er:$WS_2$ flakes of different thicknesses (Fig. 4). For a flake of thickness $\delta \approx 200$ nm, we observed the same qualitative behavior described above: A 200 mT magnetic field normal to the surface caused a reduction in overall PL intensity and a rotation of the polarization axis at 1540 nm. In contrast, when the flake thickness was increased to $\delta \approx 1$ µm, the field-induced variations diminished substantially, both in terms of emission amplitude and polarization orientation. This reduction of the effect in thicker samples suggests that part of the magnetic-field response is mediated by modifications to the local optical mode structure within the finite-thickness $WS_2$ layer, rather than arising solely from changes in the internal electronic configuration of the $Er^{3+}$ ions.

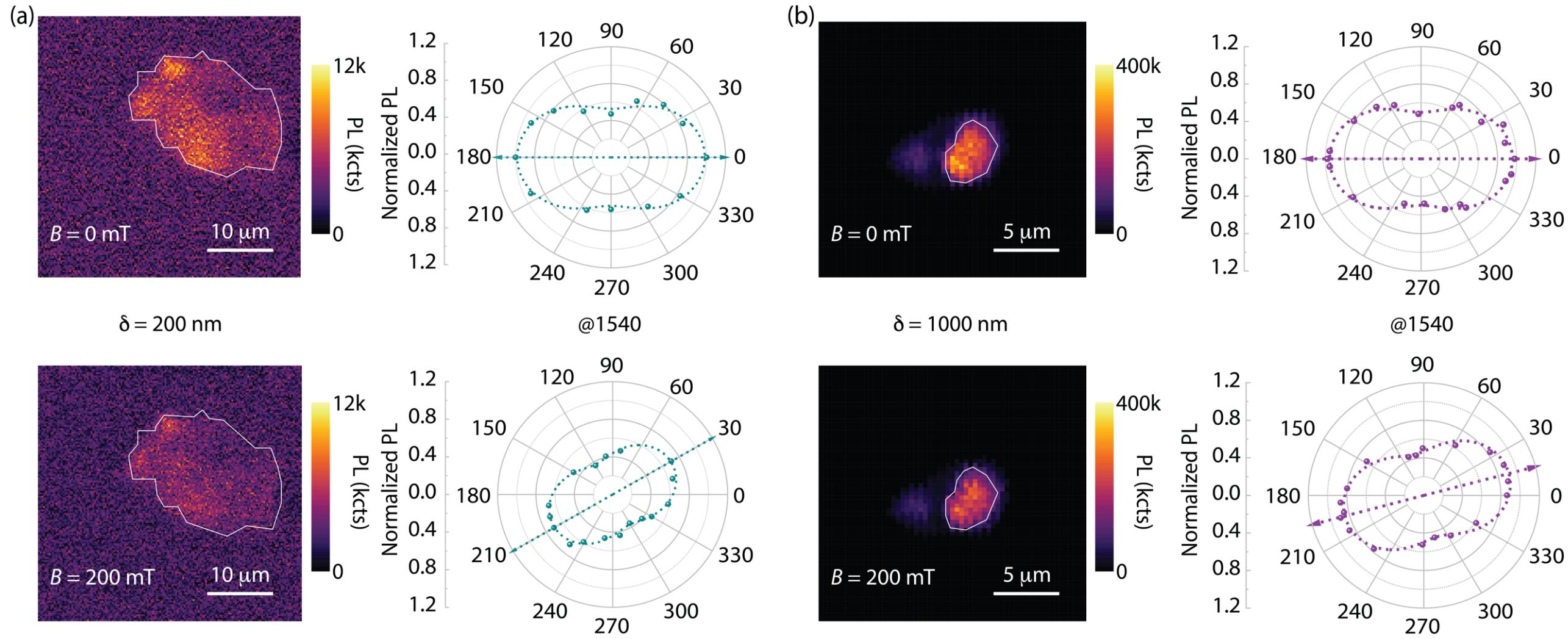

**Figure 4. Thickness-dependent response of $Er^{3+}$ emission to applied magnetic field.** (a) PL maps and corresponding polarization plots at 1540 nm from an Er:$WS_2$ flake of thickness $\delta \approx 200$ nm, recorded with and without a 200 mT magnetic field normal to the surface. The applied field induces both a reduction in overall emission and a rotation of the polarization axis, similar to that observed in Figs. 1 and 2 for a 400 nm thick flake. (b) Measurements from a thicker flake ($\delta \approx 1$ µm) under the same conditions. In this case, the field produces reduced changes in either PL amplitude or polarization pattern. The loss of field sensitivity with increasing thickness suggests a contribution from photonic effects associated with the finite optical mode structure of the flake.

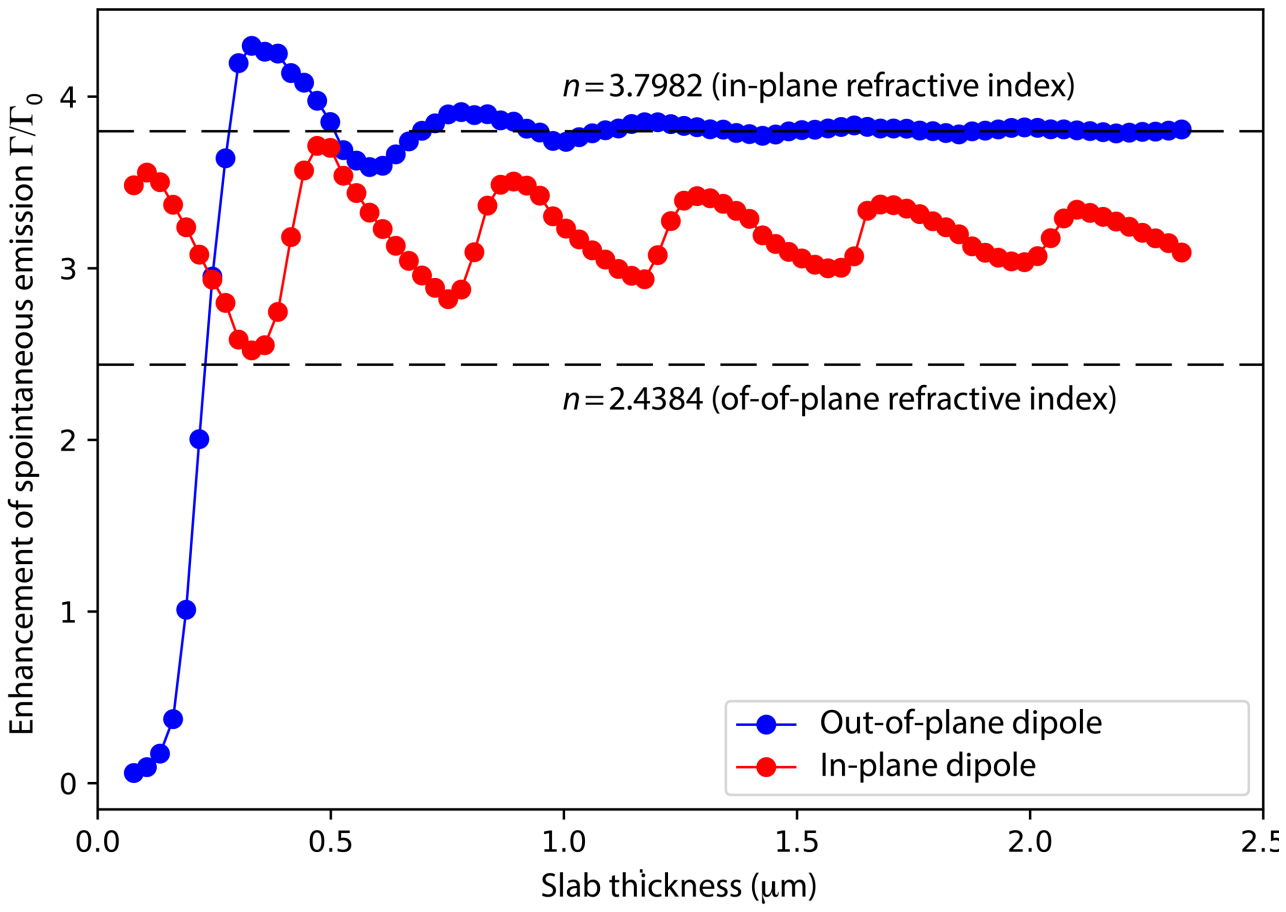

**Figure 5. Calculated spontaneous-emission enhancement for a dipole embedded in a $WS_2$ slab as a function of slab thickness.** The emission rate $\Gamma/\Gamma_0$ is shown for dipoles polarized out of plane (blue) and in plane (red). The dashed horizontal lines indicate the bulk limits corresponding to the in-plane ($n_{\parallel} = 3.7982$) and out-of-plane ($n_{\perp} = 2.4384$) refractive indices of $WS_2$. Larger changes in the emission are expected for flake thicknesses below ~400 nm.

To further explore this interpretation, we next evaluate how the LDOS evolves with flake thickness and how these variations can influence both the radiative rate and polarization characteristics of embedded emitters[13]. Figure 5 shows the calculated spontaneous-emission enhancement $\Gamma/\Gamma_0$ for dipoles embedded in a $WS_2$ slab as a function of thickness, taking into account the anisotropic refractive indices of the material. The results reveal a pronounced dependence on dipole orientation. For dipoles polarized normal to the layer, the emission rate increases rapidly with thickness and approaches the bulk limit associated with the in-plane refractive index. In contrast, dipoles polarized within the plane converge to an intermediate value reflecting contributions from both in-plane and out-of-plane dielectric responses. For thin slabs, the emission rate exhibits oscillatory behavior arising from interference between the emitter and reflections at the slab interfaces. These calculations therefore indicate that changes in dipole orientation — such as those induced by magnetic-field mixing of the CF states — can modify the coupling of the emitter to the photonic modes of the finite-thickness $WS_2$ layer, thereby contributing to the thickness-dependent variations in emission amplitude and polarization observed experimentally. A more complete understanding of this interplay between atomic and photonic effects will require systematic studies on $WS_2$ flakes of controlled and varying thickness.

## III. CONCLUSIONS AND OUTLOOK

We have shown that $Er^{3+}$ ions incorporated into $WS_2$ flakes exhibit a strongly orientation-dependent magneto-photonic response at telecom wavelengths. Using time- and polarization-resolved photoluminescence under modest magnetic fields ($\leq 0.2$ T), we observed pronounced dimming, lifetime extension, and dipole-axis rotation when the field has an out-of-plane component. Distinct responses at 1520 nm and 1540 nm further reflect differences in the underlying Stark transitions and their magnetic/electric dipole admixtures. These effects likely originate from Zeeman-induced mixing between nearly degenerate CF sublevels, which modifies the magnitude and orientation of the optical transition dipole moments. Effective model calculations of $Er^{3+}$ in monolayer $WS_2$ qualitatively reproduce these mechanisms, capturing the interplay of Stark splitting, $g$-tensor anisotropy, and dipole reorientation under external fields. Finally, the field-induced dipole reorientation may also influence how the emitters couple to the anisotropic photonic environment of the $WS_2$ flakes, potentially contributing to the observed lifetime and polarization changes.

Looking forward, engineering flakes of controlled thickness and integrating them with photonic cavities or waveguides could amplify the magneto-photonic effects observed here, and enable dynamic control of $Er^{3+}$ emission directionality, polarization, and timing. Such capabilities point to potential applications as field-tunable telecom-band quantum light sources, spin–photon interfaces, and quantum memory elements in integrated photonic platforms. Beyond control devices, the observed sensitivity of $Er^{3+}$ emission to external magnetic fields suggests an intriguing route for all-optical magnetometry. In particular, Er:$WS_2$ could serve as an optical probe of layered magnetic materials such as CrSBr[14,15], where coupling between $Er^{3+}$ TDMs and local magnetic order would provide a direct window into field-dependent spin textures with no need for microwave manipulation.

Other perturbations such as local strain can in principle produce comparable modifications of $Er^{3+}$ emission. Given the susceptibility of atomically thin $WS_2$ to mechanical deformation, strain-induced changes in CF splitting and/or dipole orientation could similarly reduce radiative coupling and extend the excited-state lifetime. Disentangling magnetic and strain effects in future studies will therefore be essential, not only to expand our understanding of the magneto-photonic response reported here but also to establish strain engineering as a complementary pathway to tune $Er^{3+}$ emission in 2D hosts.

## ACKNOWLEDGEMENTS

G.G.A. and C.A.M. acknowledge support from the Department of War, Award W911NF-25-1-0134. C.E.D. and G.I.L.M. acknowledge support from the National Science Foundation under award DMR-2237674. All authors also acknowledge access to the facilities and research infrastructure of the NSF CREST IDEALS, grant number NSF-2112550. The Flatiron Institute is a division of the Simons Foundation.

## APENDIX A: MATERIALS AND METHODS

### 1. Sample preparation

$WS_2$ flakes were mechanically exfoliated from a high-purity bulk crystal (2D Semiconductors) and transferred onto 4 × 4 $mm^2$ silicon substrates (0.5 mm thick) that had been patterned by UV photolithography to facilitate flake identification. Erbium ions were introduced by broad-area implantation at 75 keV with a fluence of $10^{14}$ $cm^{-2}$, using a 22° tilt to suppress channeling. Following implantation, the samples were annealed for 1 h at 400 °C in an argon atmosphere, with a temperature ramp of 6.3 °C/min in a tube furnace (Across International STF1200), as described previously[8].

### 2. Experimental setup

We perform optical measurements using a home-built scanning confocal microscope optimized for telecom emission. Excitation is provided either by a continuous-wave 980 nm diode laser (Optoengine, 1 W maximum output) or by a 500-mW pulsed laser diode (LaserLand SLD980500TH-A, driven by an IC Haus EVAL HB unit). For imaging and steady-state spectroscopy we use the CW source, attenuated to ~10 mW at the sample plane. For time-resolved lifetime measurements we employ the pulsed source, with an average power of ~5 mW at the sample. Both beams are coupled into single-mode fiber, collimated, and focused onto the sample with a 50×, NA = 0.8 Olympus objective mounted on a piezo stage. The beam spot diameter is ~1 μm, and sample scanning is implemented with an XYZ piezo stage (Npoint). Emission from $Er^{3+}$ in $WS_2$ is collected through the same objective, separated from the excitation path with a dichroic mirror (Thorlabs DMLP-1138), filtered with a stack of low-pass filters to suppress residual pump light, and coupled into single-mode fiber for detection by a superconducting nanowire single-photon detector (SNSPD ID281 IDQ Quantique). Photon counts are acquired with a National Instruments PCIe card under LabVIEW control. Polarization-resolved measurements are carried out by placing a linear polarizer in the collection path to analyze the emission, while a combination of a half-wave plate and linear polarizer in the excitation path allowed us to vary the incident polarization.

## APENDIX B: RADIATIVE AND NON-RADIATIVE DECAY UNDER FINITE EXCITATION

We treat the emitting $Er^{3+}$ manifold as an effective two-level system with total decay rate

$$\Gamma_\mathrm{t} = \Gamma_\mathrm{r} + \Gamma_\mathrm{n} = \frac{1}{\tau}\,, \tag{B1}$$

where "r" and "n" respectively label the radiative and non-radiative rates, and $\tau$ is the measured decay time. Under continuous excitation with power $P$, the optical pumping rate is $W = \kappa P$ where $\kappa$ is a constant that takes into account absorption cross section and overlap. Using $p_e = W/(W + \Gamma_\mathrm{t})$ to denote the steady-state excited population, we write the PL intensity as

$$I(P) = \eta \Gamma_\mathrm{r} p_e = \eta \Gamma_\mathrm{r} \frac{s}{1+s} \tag{B2}$$

where we defined $s \equiv W/\Gamma_\mathrm{t} = P/P_\mathrm{s}$ with $P_\mathrm{s} \equiv 1/\kappa\tau$ denoting the saturation power. Using a prime to indicate the presence of a magnetic field, we use Eq. (B2) to express the ratio between the measured intensities rates as

$$\frac{I(P)}{I'(P)} = \frac{\Gamma_r'}{\Gamma_\mathrm{r}} \frac{\tau'}{\tau} \frac{1+s}{1+s\tau'/\tau} \tag{B3}$$

which allows us to determine the ratio $r \equiv \Gamma_r'/\Gamma_\mathrm{r}$ between the radiative rates from the observed PL dimming, lifetime change, and normalized laser power. Furthermore, defining $x \equiv \Gamma_\mathrm{n}/\Gamma_\mathrm{r}$ and assuming the non-radiative rate remains unchanged in the presence of the field, we can recast the ratio between radiative and non-radiative rates as

$$\frac{\tau'}{\tau} = \frac{\Gamma_\mathrm{t}}{\Gamma_t'} = \frac{1+x}{r+x}\,, \tag{B4}$$

which allows us to derive $x$ in terms of the observed lifetime and calculated ratio between radiative ratios. The quantum yields then follow immediately:

$$Q \equiv \frac{\Gamma_\mathrm{r}}{\Gamma_\mathrm{t}} = \frac{1}{1+x} \quad ; \quad Q' \equiv \frac{\Gamma_r'}{\Gamma_t'} = \frac{r}{r+x}\,. \tag{B5}$$

By the same token, we can calculate the absolute radiative and non-radiative rates as

$$\Gamma_\mathrm{r} = \frac{1}{\tau(1+x)} \quad ; \quad \Gamma_\mathrm{n} = \frac{x}{\tau(1+x)}\,. \tag{B6}$$

We leverage this model in Fig. 6, where we first present the measured PL intensity as a function of excitation power for a representative $Er:WS_2$ site (Fig. 6a). The data follow the saturation curve of Eq. (B2), yielding a saturation power of ~10 mW. This calibration allows us to determine the normalized excitation parameter $s$ for any given experimental condition and thereby apply Eq. (B3) to extract the ratio $r$ between radiative rates with and without an applied magnetic field. Figures 6b and 6c extend this analysis to wide-field imaging: PL maps recorded at two different powers, with and without a 200 mT normal field, show uniform dimming across the flake. From these observations and the measured PL decay times with and without applied field (Fig. 2a in the main text), we infer a quantum yield of ~93% at zero field that decreases to ~85% under a 200 mT normal field, as presented in the main text. Using Eq. (B6), the room-temperature radiative rates with and without field are estimated to be ~90 $s^{-1}$ and ~200 $s^{-1}$, respectively; the non-radiative rate is estimated at ~15 $s^{-1}$.

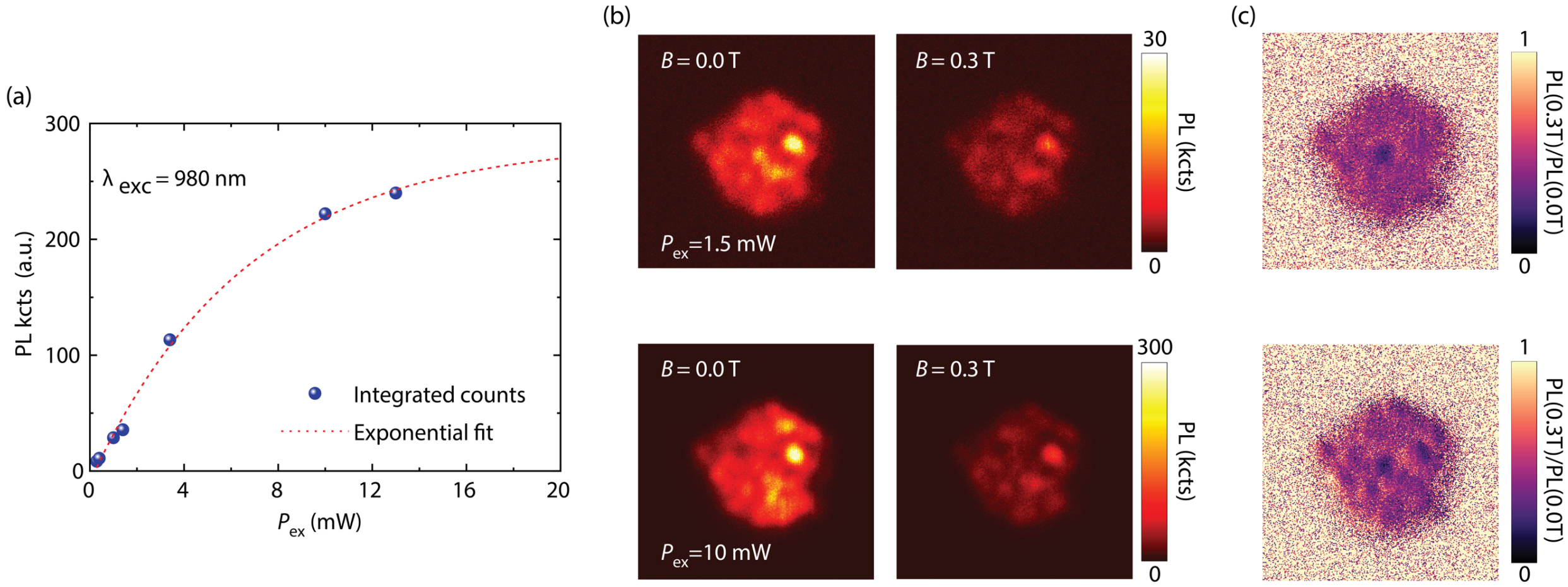

**Figure 6. Power dependence and spatial uniformity of field-induced dimming in $Er:WS_2$.** (a) PL intensity as a function of excitation power for a representative $Er:WS_2$ site; the curve follows the saturation behavior described by Eq. (B1), yielding a saturation power of approximately 10 mW. (b) Wide-field images of a flake recorded at two different powers, with and without a 200 mT magnetic field pointing normal to the surface. (c) Pixel-by-pixel ratio of the PL under field and zero-field conditions for the two cases in (b) revealing spatially uniform dimming across the flake. Working below saturation, the fractional dimming is nearly power insensitive.

## APENDIX C: NUMERICAL MODELING

### 1. First-principles calculations

The relaxed atomic configuration and ground-state electronic structure of the $\mathrm{Er}_\mathrm{W}^-$ defect in monolayer $WS_2$ are obtained using DFT. With this as starting point, we construct a semi-empirical effective model for the 4*f* states of the $\mathrm{Er_W}$ defect based on quantum embedding[8]. This approach allows us to describe the correlated $Er^{3+}$ states in a many-body footing, while capturing important host-induced mixing and hybridization. The spherically-symmetric part of our effective model takes the form

$$\hat{\mathcal{H}}_{\mathrm{eff}}^{\mathrm{sph}} = \frac{1}{2}\sum_{\langle ijkl \rangle} U_{ijkl}\, c_i^\dagger c_j^\dagger c_l c_k + \lambda \hat{\mathbf{L}} \cdot \hat{\mathbf{S}}\,, \tag{C1}$$

where $c_i^\dagger$ $(c)$ are creation (annihilation) operators and $i, j, k, l$ label Er $4f$ states. $U_{ijkl}$ represent screened Coulomb matrix elements, for which we assume a spherically symmetric Slater form with Hund's coupling parameter $J$ chosen such that the many-body spectrum of an isolated $Er^{3+}$ best matches that from experiments[6,11,16,17]. An analogous procedure is used to obtain the spin-orbit interaction strength $\lambda$. We construct the spin-orbit operator in the basis of spherical harmonics ordered by increasing angular momentum quantum number, using $\hat{S}_{x,y,z} = \frac{1}{2}\sum_{i\sigma\sigma'} c_{i\sigma}^\dagger \tau_{\sigma\sigma'}^{x,y,z} c_{i\sigma'}$ (where $\tau^{x,y,z}$ are Pauli matrices) and $\hat{L}_{z,+,-} = \sum_{ii'\sigma} c_{i\sigma}^\dagger L_{ii'}^{z,+,-} c_{i'\sigma}$, where $L_{ii'}^z = i\delta_{i,i'}, L_{ii'}^+ = \delta_{i,i'+1}\sqrt{l(l+1) - i'(i'+1)}$ and $L_{ii'}^- = \delta_{i,i'-1}\sqrt{l(l+1) - i'(i'-1)}$. We translate this into the cubic $Y_{lm}$ basis via

$$T = \frac{1}{\sqrt{2}}\begin{bmatrix} 1 & 0 & 0 & 0 & 0 & 0 & -1 \\ 0 & 1 & 0 & 0 & 0 & 1 & 0 \\ 0 & 0 & 1 & 0 & -1 & 0 & 0 \\ 0 & 0 & 0 & \sqrt{2} & 0 & 0 & 0 \\ 0 & 0 & i & 0 & i & 0 & 0 \\ 0 & i & 0 & 0 & 0 & -i & 0 \\ i & 0 & 0 & 0 & 0 & 0 & i \end{bmatrix}.$$

Only the quadratic part of the spin-orbit operator is retained.

Spherical symmetry is broken via the CF interactions

$$\hat{\mathcal{H}}_{\mathrm{eff}} = \hat{\mathcal{H}}_{\mathrm{eff}}^{\mathrm{sph}} + \hat{\mathcal{H}}_{\mathrm{CF}}\,, \tag{C2}$$

which we describe via

$$\hat{\mathcal{H}}_{\mathrm{CF}} = C\sum_{\langle ij \rangle} t_{ij} c_i^\dagger c_j + \sum_{a,b} K_{ab}\, \hat{\mathbf{J}}_a \hat{\mathbf{J}}_b\ , \tag{C3}$$

with $\hat{\mathbf{J}}_a = \hat{\mathbf{L}}_a + \hat{\mathbf{S}}_a$ and $a, b \in \{x, y, z\}$, where only the quadratic part of the second term is retained. Namely, the CF environment contains the hopping matrix elements $(t_{ij})$ and a single-particle MCA-like "correction" via the $K_{ab}$ tensor. We add such MCA term to the *ab-initio* CF in view of potentially missing contributions (e.g., spin-orbit coupling with $WS_2$, CF from other layers, etc.) that are likely to refine the orientation of the magnetic moments in the absence of external fields and their energy landscapes. The $t_{ij}$ matrix elements are obtained by disentangling and Wannierizing the Er $4f$ manifold of states from non-magnetic DFT calculations[18]. The direct output from DFT is symmetrized under the $D_{3h}$ point group, which is the expected site symmetry of a perfectly substitutional $Er^{3+}$ defect in $WS_2$. We find the obtained *ab-initio* CF to yield overestimated CF splitting with respect to $Er^{3+}$ dopants in numerous other solids[6,11], likely originating from the artificially enhanced mixing/hybridization in DFT. Scaling this term by $C = 0.25$ yields reasonable CF splitting. Motivated by the experimental angular dependence of the magnetic-field induced effects, we assume a $D_{3h}$-symmetric form of the MCA tensor,

$$K_{ab} = K\begin{bmatrix} 1 & 0 & 0 \\ 0 & 1 & 0 \\ 0 & 0 & 0 \end{bmatrix}, \tag{C4}$$

which in our basis reduces the MCA CF contribution in Eq. (C3) to $\sum_{a,b} K_{ab}\, \hat{\mathbf{J}}_a \hat{\mathbf{J}}_b = K\,(\hat{\mathbf{J}}^2 - \hat{\mathbf{J}}_z^2)$. The parameter $K$ is determined using the following procedure. First, we employ individual relativistic DFT calculations where the magnetic moment of the $\mathrm{Er_W}$ defect is constrained along different Cartesian directions (e.g., *x*, *y*, *z*, *xy*, *xz*, *yz*). From these results, we find the largest energy barrier $\Delta E_{\mathrm{MCA}} = 6.7$ meV. Provided the aligned magnetic moment has $\langle J_i^2 \rangle \approx J(J+1)$ along its preferred axis, i.e., the constrained moment is purely axial, we define $\Delta E_{\mathrm{MCA}} \equiv KJ(J+1)$. Lastly, assuming our relativistic DFT calculations correctly predict the $^4\mathrm{I}_{15/2}$ term as the $Er^{3+}$ ground-state (strongly supported by the obtained single-particle electronic structure), we have $K = \frac{\Delta E_{\mathrm{MCA}}}{J(J+1)} \approx 0.11$ meV. With this parametrization, we find qualitative agreement between the MCA derived directly from non-relativistic DFT calculations and our MCA-corrected effective CF Hamiltonian. Additionally, we find equivalent results (i.e., state splitting/degeneracies and **B**-induced effects) with slightly different $D_{3h}$-symmetric forms of the tensor in Eq. (C4), provided a slightly tuned $K$ parameter. We point out that the magnetic and spin-orbit interactions neglected at the DFT level can change the ordering of our many-body states. Thus, the states featuring easy-plane MCA within our effective model could be the lowest CF states thermally populated in experiments. Such scenario could explain why all emission peaks observed in experiments behave the same way under the influence of magnetic fields.

We access the optical transitions of the $\mathrm{Er}_\mathrm{W}^-$ defect by evaluating the dipole matrix elements across the many-body spectrum using

$$\boldsymbol{\mu}_{ij} = e\,\langle \Psi_i | \hat{\mathbf{d}} | \Psi_j \rangle\,, \tag{C5}$$

where $\Psi_{i,j}$ denote many-body wave functions of states $i$

and $j$, and $\hat{\mathbf{d}}$ represents the many-body dipole operator. Eq. (C5) involves many-body ground and excited states of the $\mathrm{Er}_\mathrm{W}^-$ defect, including important correlation effects neglected at the DFT level. Usually, the $4f \leftrightarrow 4f$ transitions in $Er^{3+}$ around 1550 nm have strong magnetic character, though forced-ED (and/or electric quadruple) components can have comparable contributions to the total amplitudes depending on the host and site-symmetry[2,3,6,11]. Thus, we consider $\hat{\mathbf{d}} = \hat{\mathbf{r}} + \hat{\mathbf{m}}$, with

$$\hat{\mathbf{r}} = \sum_{nm} \mathbf{r}_{nm}^{\mathrm{wan}}(\mathbf{0}) c_n^\dagger c_m \tag{C6}$$

and

$$\hat{\mathbf{m}} = \frac{\hbar}{m_\mathrm{e} c} (\hat{\mathbf{L}} + g_\mathrm{e} \hat{\mathbf{S}}) \tag{C7}$$

describing the electric and magnetic components, respectively. $\mathbf{r}_{nm}^{\mathrm{wan}}(\mathbf{0})$ represent matrix elements of the position operator between Wannier functions $m$ and $n$, given by

$$\mathbf{r}_{nm}^{\mathrm{wan}}(\mathbf{R}) = \langle n\mathbf{0} | \mathbf{r} | m\mathbf{R} \rangle. \tag{C8}$$

These matrix elements are also symmetrized according to the selection rules under $D_{3h}$ symmetry. The electron $g$-factor is taken as $g_\mathrm{e}$= 2.002319. Lastly, decay rates for select many-body transitions near 1550 nm (0.8 eV) are accessed via the Einstein coefficients $A_{ij} = \frac{n_\mathrm{r} \omega_{ij}^3}{3\pi\varepsilon_0 \hbar c^3} |\boldsymbol{\mu}_{ij}|^2$, using the matrix elements obtained from Eq. (C5) and 3.798 (2.438) for the in-plane (out-of-plane) refractive index of bulk $WS_2$ ($n_\mathrm{r}$) at 1550 nm[19].

Lastly, we calculate adiabatic MCA energy surfaces as follows. First, we project our many-body Hamiltonian into a select manifold (either $^4I_{15/2}$ or $^4I_{13/2}$) and perform a Stevens coefficient extraction (least-squares fit) including all Stevens operators allowed under $D_{3h}$ symmetry. This procedure results in a set of Stevens coefficients that best describes a given fixed-$J$ manifold of our effective cubic-$Y_{lm}$ CF+SOC Hamiltonian in the $\hat{O}_q^k$ basis. With these at hand, we construct a 3D grid, at which individual point we (i) rotate the Stevens CF Hamiltonian, (ii) project onto a specific many-body doublet, (iii) extract the lowest eigenvalue. Formally, $E_{\mathrm{MCA}}(\theta,\phi) = \min \mathrm{eig}\ [PU(\theta,\phi) H_{CF}^{\mathrm{S}} U^\dagger(\theta,\phi) P]$, where $P$ is the projector that takes the fixed-$J$ Stevens CF Hamiltonian $H_{CF}^{\mathrm{S}}$ obtained from the least-squares fit into a given Kramers doublet subspace, and $U(\theta,\phi)$ are unitary transformations that rotate $H_{CF}^{\mathrm{S}}$ around the 3D grid.

All DFT calculations are performed using the VASP code[20], with exchange-correlation interactions described via the semi-local generalized-gradient approximation (GGA) parameterized by Perdew, Burke and Ernzerhof (PBE)[21]. Core electrons are treated via the projector-augmented wave (PAW) pseudopotentials[22], with erbium $4f$ electrons included in the valence. A kinetic energy cutoff of 700 eV is chosen for the plane-wave basis. We model the defect by using a 7×7×1-supercell containing a single $\mathrm{Er}_\mathrm{W}$ substitution, with a 20-Å vacuum region to avoid interlayer interactions. All calculations are employed at the Γ-point only, with tight convergence criteria for forces (0.001 eV/Å) and electronic iterations ($10^{-8}$) to ensure well-converged structures and wave functions. The ground-state atomic structure of the $\mathrm{Er}_\mathrm{W}$ defect is obtained via spin-polarized DFT calculations, while all other subsequent (e.g., Wannier) calculations are performed without spin-polarization (non-magnetic). All Wannier function (and related) calculations are employed using the Wannier90 package as interfaced with VASP[18]. Consistent with the $Er^{3+}$ configuration, we take the $4f$ manifold (7 orbitals in the non-magnetic case) as the active space for our quantum embedding calculations, containing 11 electrons. The disentanglement energy window used for Wannier90[23] is [-3.15, 0.50] eV around the Fermi level of the full defect to capture important hybridization effects between erbium, bulk $WS_2$ states, and neighboring dangling bonds. The construction and diagonalization of our correlated model is employed using the Toolbox for Research on Interacting Quantum Systems (TRIQS) software library[24].

### 2. Modeling photonic effects

The spontaneous emission enhancement factor was computed using the finite-difference time-domain (FDTD) method as implemented in the open-source package Meep[25]. An out-of-plane (in-plane) polarized electric dipole source was placed at the center of a $WS_2$ slab (surrounded by air) with anisotropic permittivity (in-plane index $n = 3.7982$, out-of-plane index $n = 2.4384$[26]) at a free-space wavelength of $\lambda = 1.55$ µm. The enhancement factor was obtained by normalizing the LDOS in the presence of the slab to the LDOS of the same dipole in free space. We simulate various slab thicknesses from $0.05\lambda$ to $1.5\lambda$. Perfectly matched layers were applied at all boundaries to absorb outgoing radiation.